\documentclass{elsart}
\usepackage{icarus}

\usepackage{natbib}

\bibpunct{(}{)}{;}{a}{,}{,}

\usepackage[dvips]{graphicx}
\newcommand{\BibTeX}{ \textrm{B\kern-.05em\textsc{i\kern-.025em b}\kern-.08em
    T\kern-.1667em\lower.7ex\hbox{E}\kern-.125emX} }

\begin{document}

\begin{frontmatter}



\title{Appearance of Saturn's F ring azimuthal channels for the anti-alignment configuration between the ring and Prometheus}


\author[Carlos]{Carlos E. Chavez},

\address[Carlos]{Instituto de Astronomia UNAM sede Ensenada, Km. 103 carretera Tijuana-Ensenada, c.p. 22860, Ensenada, Baja California, Mexico}

%% This copyright statement isn't required at any stage by the Icarus
%% Editorial Office or Elsevier.  However, until you sign over the
%% copyright to Elsevier prior to publication (or negotiate with them
%% about copyright), you own the copyright to anything you create.
%% Just to keep things unambiguous, always include a copyright statement
%% or explicitly dedicate your work to the public domain.

%% ----- ELSEVIER STUFF -----
%% The commands below up to the \end{frontmatter} are commented out
%% so that we can do some Icarus-required formatting on the second and
%% third pages that is not required later on by Elsevier.  So when
%% your paper gets accepted, and you are ready to start dealing with
%% Elsevier, copy your abstract and keywords up here, uncomment these
%% lines, and comment out the ICARUS STUFF below.
%% 
%% Alternately, you might just want to move these abstract, keyword,
%% and end frontmatter commands down, and comment out the ICARUS STUFF
%% commands.  It doesn't matter.

% \begin{abstract}
% % Text of abstract
% 
% \end{abstract}
% 
% \begin{keyword}
% % keywords here, in the form: keyword \sep keyword
% 
% 
% % PACS codes here, in the form: \PACS code \sep code
% 
% \end{keyword}

%% ----- END ELSEVIER STUFF -----

\end{frontmatter}

%% ----- ICARUS STUFF -----
%% Some formatting on the first, second, and third pages are required
%% by the Icarus Editorial Office that are not required by Elsevier.
%% This section contains those things.  When you are ready to transition
%% to ``Elsevier'' mode, copy your abstract and keywords up into
%% the ELSEVIER STUFF section, and then you can just delete everything
%% in this section.

%% We need to list the number of manuscript pages, figures, and tables. 
%%
%% Rather than manually count these things out, we'll use a little
%% trick here from Paul.  All you have to do is place three \label{}
%% tags on the last page, the last table, and the last figure, that
%% way these values are automatically updated (as long as you remember
%% to move the lasttable and lastfig labels when you add or remove
%% tables and figures).

\begin{flushleft}
\vspace{1cm}
Number of pages: \pageref{lastpage} \\
Number of tables: \ref{lasttable}\\
Number of figures: \ref{lastfig}\\
\end{flushleft}

%% Don't worry about finding the various last* tags and deleting them
%% when you go to ``Elsevier'' mode if you don't want to, they should be
%% silently ignored.

%% The second page should indicate a proposed running head of not more 
%% than 55 characters, and the name and address to which editorial 
%% correspondence and proofs should be directed.  The pagetwo 
%% environment that icarus.sty provides will make page two for you,
%% just give the running head as an argument to the environment, and
%% then your correspondence address inside.
\begin{pagetwo}{Appearance of F ring azimuthal channels at anti-alignment }
%                        1         2         3         4         5
%               1234567890123456789012345678901234567890123456789012345

Carlos E. Chavez \\
Instituto de Astronomia UNAM sede Ensenada\\
Km. 103 carretera Tijuana-Ensenada\\
Ensenada, Baja California, c.p. 22860, Mexico. \\
\\
Email: carlosepech@astrosen.unam.mx\\
Phone: (0052) 6461744593 ext. 326 \\
Fax: (0052) 6461744607

\end{pagetwo}

\begin{abstract}
In this article we explore the aspect of the F ring with respect to the anti-alignment configuration between the ring and Prometheus.  We focus our attention on the shape of the F ring's azimuthal channels which were first reported by~\cite{Porco_Science_2005} and numerically explored by~\cite{Nature_2005}, who found excellent agreement between Cassini's ISS reprojected images and their numerical model via a direct comparison.
We find that for anti-alignment the channels are wider and go deeper inside the ring material. From our numerical model we find a new feature, an island in the middle of the channel. This island is made up of the particles that have been perturbed the most by Prometheus and only appears when this satellite is close to apoapsis. In addition, plots of the anti-alignment configuration for different orbital stages of Prometheus are obtained and discussed here.

\end{abstract}

% %% Keywords should appear after the abstract. 

%% ----- END ICARUS STUFF -----

%main text
\section{Introduction}
Since the azimuthal channels were first discovered with the Cassini spacecraft in July 2004 (\cite{Porco_Science_2005}), these features have captured the imagination of scientist involved in the Cassini mision. These channels have appeared 28 times in the imaging diary and appeared in one press release in the ciclops website (\texttt{http://www.ciclops.org}) as counted on June 1, 2008. 
Numerical simulations of the F ring have shown the possibility that azimuthal channels may form (\cite{Showalter_Burns};~\cite{Winter_2000}). But it was not until the Cassini spacecraft arrived, that this was confirmed (\cite{Porco_et_al_2006}). The available initial conditions obtained from these images have permitted us to make a direct comparison between Cassini's ISS images and the numerical simulations (\cite{Nature_2005}) for the first time.
The dynamical complexity of the channels is evident when we see how their aspect changes over time, as a function of Prometheus orbit (\cite{Murray_Winter_1996},~\cite{Winter_2000} and~\cite{Nature_2005}). The observed changing angle is believed to be due to Keplerian shear (\cite{Nature_2005}). 
Here we study the anti-alignment configuration between the F ring and Prometheus and the consequences for the azimuthal channels using numerical full integrations.

\section{Methods}
Possible complications that our treatment ignores, include (i) The gravitational influence of other satellites (Prometheus is without doubt the origin of the channels, as found by ~\cite{Nature_2005}). (ii) The oblateness of Saturn and therefore its precessional effect on the orbits of the ring particles and Prometheus (mutual precession is only $0.057^{\circ}d^{-1}$, for the time scales involved here and it can be neglected). (iii) The effect of relatively large objects within the F ring core and therefore self-gravity. (iv) Collisions between F ring particles (this is a standard assumption when modeling the F ring as it was supposed in ~\cite{Nature_2005}, ~\cite{Winter_2000}, ~\cite{Murray_1997}, etc.).

The approximate date for the anti-alignment configuration has been calculated using recently published ephemerides.~\cite{Jacobson_2008} Table 5 was used for Prometheus and the solution of ~\cite{Bosh_2002} for the F ring. 
We proceeded as follows: First we took an arbitrary initial orbital configuration, here we used the Cassini's Saturn Orbital Insertion (hereafter SOI) date as the initial date. This date corresponds to UT 04:48:00 on July 1,  2004. For that orbital configuration, we calculated the minimum distance between Prometheus and the F ring by dividing the orbit in 1000 points (which gives us an error in the calculation of the order of $0.36^{\circ}$). Then we use precessing ellipses to evolve the orbits, with that we obtain a new orbital configuration. Then we repeat the calculation of the minimum distance, and so on.
The minimum distance between Prometheus and the F ring core is calculated over a time span of 44 years with a time step of 1 day.

The values of the orbital elements obtained by this method are shown in Table~\ref{I10533table}. 
For convenience we used the orbital elements in the reference frame in which $\widetilde{\omega}=\Omega=0$ for Prometheus. This means that in this reference frame $\widetilde{\omega}=179.8648^{\circ}$ and $\Omega=271.3077^{\circ}$ for the ring.

\begin{figure}
\begin{center}
\includegraphics[width=12cm,height=10cm]{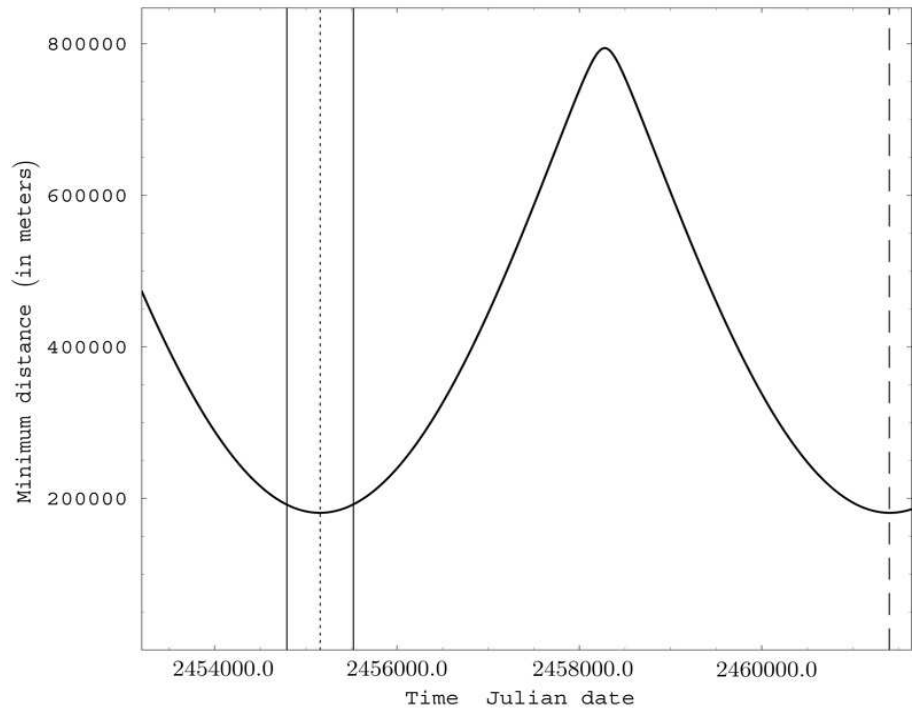}
\caption[Minimum distance between Prometheus and the core as a function of time.]{
The minimum distance between Prometheus and the F ring core is shown here (in meters) versus time (in Julian Date). The dotted and dashed vertical lines correspond to the next anti-alignment (November 21, 2009) and the one after it (December 24, 2026), respectively.
The two vertical solid lines, one at the left and the one at the right of the dotted line, correspond to a date one year earlier and after the first anti-alignment, respectively. 
}
\label{I10533fig1}
\end{center}
\end{figure}

Figure \ref{I10533fig1} shows how the minimum distance between Prometheus and the F ring core changes over time. Starting from epoch JD 2453187.7 (corresponding to UT 04:48:00 on July 1,  2004 which is the SOI date) and finishing in epoch JD 2467787.7 (corresponding to UT 08:17:49.3 on August 27, 2027). Then the first minimum on the plot corresponds to the approximated date of the next anti-alignment event, shown in Fig. \ref{I10533fig1} as a dotted line. This date has been found to be epoch JD 2455156.5 (corresponding to UT 4:54:24 on November 21, 2009) and the corresponding minimum distance is 180.9 km. Since there are no sharp or abrupt changes in the plot (the shape is very smooth when approaching to minimum), the systems shows a soft transition to the anti-alignment configuration. 
This means that the shape of the channels should not change substantially around anti-alignment.
In Fig. \ref{I10533fig1} we are showing dates one year before and one year after the encounter, shown here as two solid vertical lines (one at the left and one at the right of the minimum, respectively).
In Fig. \ref{I10533fig1}, the difference in distance between the minimum distance at anti-alignment and the line at the left is 10.8 km, and for the one at the right is 11.3 km. These distances are very small compared to 472.5 km, which is the minimum distance at SOI. These dates are relevant since the Cassini spacecraft extended mission will end in July 1, 2010.
  
Once per orbit during this anti-alignment Prometheus will be at apoapsis and the ring particles will be at periapsis. 
This configuration is studied here numerically, and the possible appearance of the F ring for such a configuration is predicted. 
Much progress has been made in understanding the nature of the F ring's strands', it was found that these strands have spiral shapes (\cite{Charnoz_2005}). More recently, it was found that small moonlets could be the origin of the strands (\cite{Murray_2008}). Despite all these advances in understanding, it is still not possible to predict the number and location of strands in the F ring at a given time and at a given longitude.Therefore, it was decided not to include strands in the numerical simulation, except the core and the background sheet (we are using the names of each feature as it appears in~\cite{Nature_2005}).

\section{Results}

\begin{figure}
\begin{center}
\includegraphics[width=16cm,height=14.15cm]{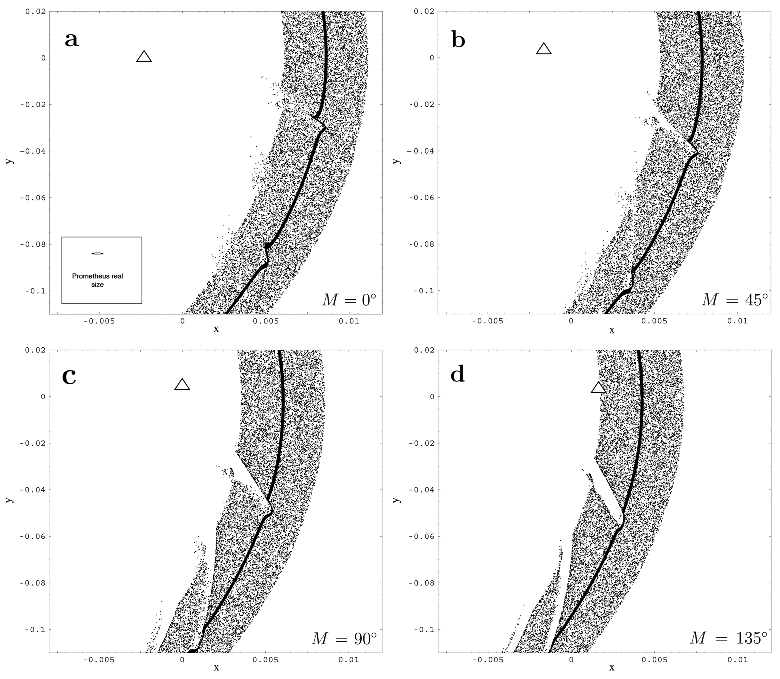}
\caption[First four snapshots]{
First four snapshots of the encounter between Prometheus and the F ring. Prometheus is represented as a triangle ($ \triangle $) and it is used to denote the position of the center of Prometheus, the true-scale of Prometheus is shown, as a elongated ellipse, in the left-bottom corner of Fig. \ref{I10533fig2}$\mathbf{a}$.
These plots are in the reference frame that rotates with Prometheus mean motion. The mean anomaly $M$ for each of the plots is shown in the right bottom part of the plot. We observe that the ring is composed by two components: The core strand (at the center of the ring) and the background sheet (the less dense component that surrounds the core).

}
\label{I10533fig2}
\end{center}
\end{figure}

\begin{figure}
\begin{center}
\includegraphics[width=16cm,height=14.15cm]{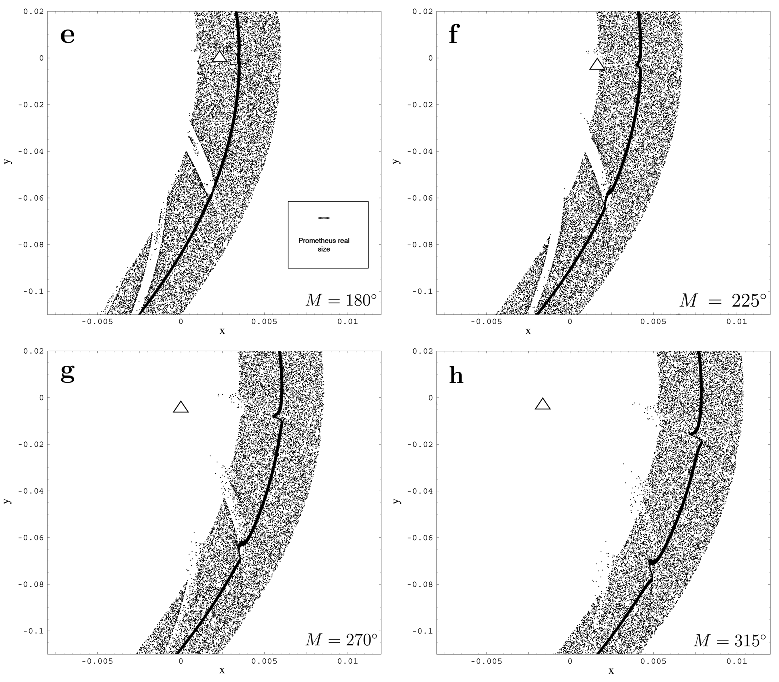}
\caption[Last four snapshots]{
Last four snapshots of the encounter between Prometheus and the F ring. Prometheus is represented as a triangle ($\triangle$) and it is used to denote the position of the center of Prometheus, the true-scale of Prometheus is shown, as a elongated ellipse, in the left-bottom corner of Fig.\ref{I10533fig3}$\mathbf{e}$.
These plots are in the reference frame that rotates with Prometheus mean motion. The mean anomaly $M$ for each of the plots is shown in the right bottom part of the plot. We observe that the ring is composed by two components: The core strand (at the center of the ring) and the background sheet (the less dense component that surrounds the core).
}
\label{I10533fig3}
\end{center}
\end{figure}

An integration of the full equations was performed for the anti-alignment configuration, with 100,000 test particles (50,000 on each strand) using the Runge-Kutta-Nystr\"om RKN 12(10) 17M of~\cite{Brankin_1989} (using the higher-order version). Numerical integrations were performed in a Saturnocentric reference frame, in which the $x$$-$$y$ plane corresponds to the equatorial plane of Saturn. As is conventionally assumed, the direction of the $x$$-$axis points to the ascending node of Saturn's equator on the Earth's mean equator of J2000 (2451545.0 JED). 
The results of the integration for the anti-alignment configuration are shown in Fig. \ref{I10533fig2} and \ref{I10533fig3} (these plots can be compared with those obtained in~\cite{Nature_2005} Fig. 2 $\mathbf{a}$ to $\mathbf{f}$). The perturbation is more extreme in this case; the streamers from previous
encounters are present and are more prominent than the ~\cite{Nature_2005} plots.
The positions of Prometheus and all the test particles are shown in the rotating reference frame
with Prometheus' mean motion.

We show in Fig. \ref{I10533fig2} and \ref{I10533fig3}  eight snapshots of the system for different orbital stages. The total time between the first plot and the last is one orbital period of Prometheus (that is 14.7462 hours), which is the natural period of time to be used since Prometheus is the cause of these features. 
The azimuthal channels that have been formed in previous Prometheus' orbits are clearly visible here. The structure and physical appearance of the channels are clearly time-dependent. These distortions are located in the inner part of the ring since Prometheus has a stronger influence here.

Figure \ref{I10533fig2} $\mathbf{a}$ (mean anomaly $M=0^{\circ}$) shows a ring with no channels in it, the streamers of material from previous passages are visible and the perturbations in the core strand are at their maximum extension, these streamers were first reported in ~\cite{Nature_2005}. Figure~\ref{I10533fig2} $\mathbf{b}$ ($M=45^{\circ}$) shows how channels start to open again, the streamers of material are still there but much smaller than in the previous plot and the perturbations in the core strand are present. Figure~\ref{I10533fig2} $\mathbf{c}$ ($M=90^{\circ}$) shows the channels, where the streamers of material have disappeared and the perturbations in the core strand have decreased substantially. Figure~\ref{I10533fig2} $\mathbf{d}$ ($M=135^{\circ}$) still shows the channels, Prometheus is getting embedded into the ring material but the perturbations in the core strand are barely visible at this point. Figure \ref{I10533fig3} $\mathbf{e} $ ($M=180^{\circ}$) shows the channels at their maximum width, Prometheus is at its closest approach to the ring's core and the perturbations in the core strand are not visible. Figure \ref{I10533fig3} $\mathbf{f}$ ($M=225^{\circ}$) shows the channels, the streamers of material have now appeared again, the perturbations in the core strand are present too, and Prometheus is still embedded in the ring material. Finally, Fig. \ref{I10533fig3} $\mathbf{g}$ ($M=270^{\circ}$) shows a configuration very similar to Fig. \ref{I10533fig3} $\mathbf{h}$ ($M=315^{\circ}$), except that in Fig. \ref{I10533fig3} $\mathbf{g}$ the channels are still present.

\begin{figure}
\begin{center}
\includegraphics[width=11cm,height=10cm]{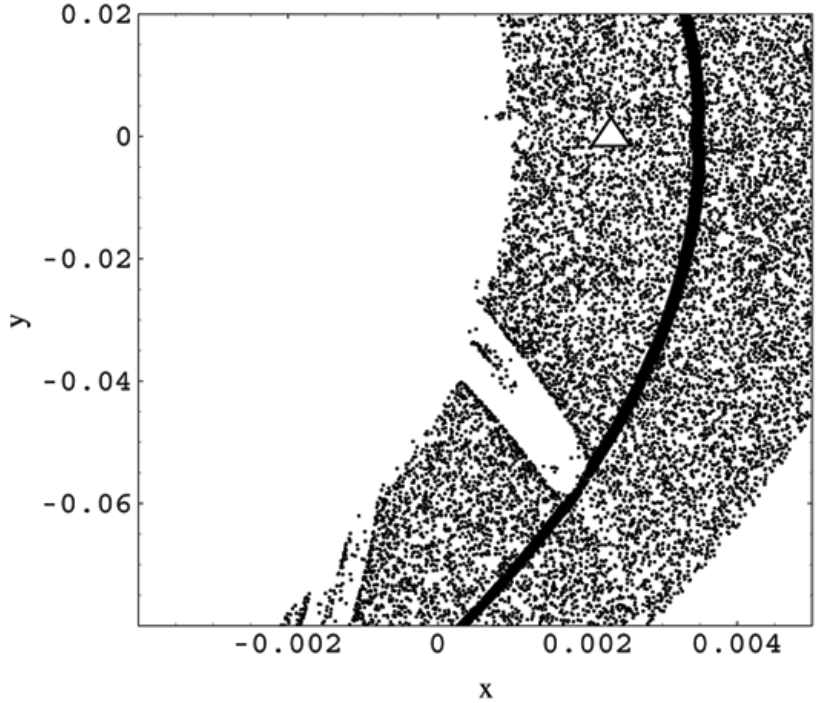}
\caption[Island of particles in the middle of the channel]{
Enlarged view of Fig. \ref{I10533fig3} $\mathbf{e}$. The mean anomaly of Prometheus is $M=180^{\circ}$. 
In this plot it is now clearer that this island is an accumulation of perturbed particles. This island appears only when Prometheus is close to apoasis (and the F ring to periapsis). 
}
\label{I10533fig4}
\end{center}
\end{figure}

It is important to notice that when Prometheus is close to apoapsis a new feature is visible in the middle of the channel for 
this configuration, as seen in Fig. \ref{I10533fig3} $\mathbf{e} $. 
In Fig. \ref{I10533fig4} we are zooming on the region in which this new island of test particles appears. From our numerical results the island is visible when Prometheus is between the mean anomaly $M=144.0^{\circ}$ to $M=201.0^{\circ}$ (that is for $\sim60.0^{\circ}$ or 2.4577 hours).

In both Figs. \ref{I10533fig2} and \ref{I10533fig3} a triangle is used to show the position of Prometheus' center, the true-scale of Prometheus is shown, as a elongated ellipse, in the left-bottom corner of Figs. \ref{I10533fig2}$\mathbf{a}$ and \ref{I10533fig3}$\mathbf{e}$.

The code utilized is able to monitor the number of collisions of ring particles with a spherical Prometheus, 50 km in radius ( the same value that was reported in~\cite{Nature_2005}) and we found that only 0.6\% of the total number of test particles included in the background sheet collided with Prometheus per orbit (and none of the particles of the core strand). The relevance of this result is that it finally rules out collisions with Prometheus as the origin of the azimuthal channels.
Here we are including a section of the ring that covers $24.0^{\circ}$ in mean longitude (from $\lambda=-7^{\circ}$ to $\lambda=17.0^{\circ}$). This longitude span was chosen because Prometheus would then encounter the F ring five times in the course of the integration. From here, and using the geometry of the ellipse, it is possible to find the area of this section of the ring, we find $A_{sect}=4.07045\times10^{7}$ $km^{2}$, which means a surface density for the ring of $\sigma=0.00122836$ particles per $km^{2}$. The total area of the ring is given by $A_{total}=6.1673\times10^{8}$ $km^{2}$. That means that the percentage of the total number of particles removed, for the whole ring, is 0.038\% per orbit, if we assume that this collisional rate continues steadily, and that the particles that collide with Prometheus stick to its surface, then all the particles of the background sheet should be removed within $\sim4.41$ years. This is just a lower value since this percentage only is this high when Prometheus is at anti-alignemnt. As reported in~\cite{Nature_2005} there were not detected collisions for the SOI configuration (which is only 5 years apart from anti-alignment). More research needs to be done to find better estimates. A detailed discussion about the collisions and the physics of the collisions will be published soon Chavez (2009) in preparation.

\begin{figure}
\begin{center}
\includegraphics[width=12cm,height=6cm]{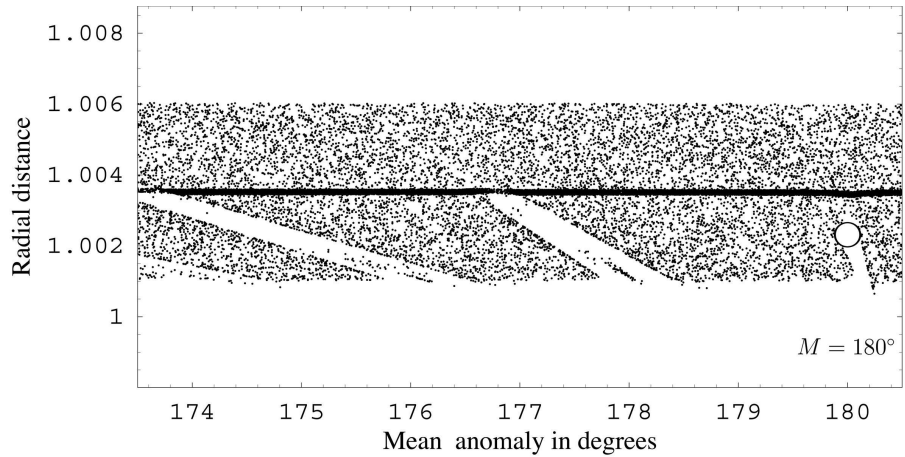}
\caption[Possible appearance of the F ring for the anti-alignment configuration with Prometheus mean anomaly $M=180^{\circ}$.]
{Plot for the anti-alignment configuration with Prometheus' mean anomaly $M=180^{\circ}$. Prometheus, as can be seen in the plot, is at its deepest incursion into the ring. Notice the prominent channels. The islands in the middle of the channels are clearly visible too.
}
\label{I10533fig5}
\end{center}
\end{figure}

\begin{figure}
\begin{center}
\includegraphics[width=12cm,height=6cm]{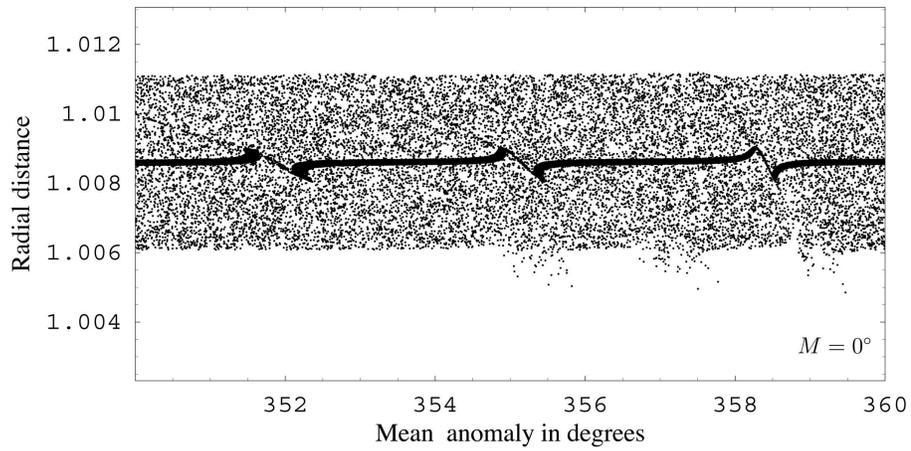}
\caption[Possible appearance of the F ring for the anti-alignment configuration with $M=0^{\circ}$.]{
Numerical model for the anti-alignment configuration, the mean anomaly of Prometheus is $M=0^{\circ}$. This is the possible
appearance of the F ring for this configuration, Prometheus is at periapsis and the F ring is at apoapsis. 
As can be noticed the streamers are pronounced. Here Prometheus is too far away from the ring to be visible in the plot.
}
\label{I10533fig6}
\label{lastfig}
\end{center}
\end{figure}

We now obtain plots of radial distance versus longitude for this configuration in order to have a better idea of the appearance of the ring for this configuration.
Fig. \ref{I10533fig5} shows the result of our numerical model for Prometheus' at amean anomaly of $M=180.0^{\circ}$. Here Prometheus is exactly at apoapsis and the F ring at periapsis, indicating that they are at their closest distance possible.
As expected, the channels, are more prominent and go deeper into the ring material for this configuration, than the ones reported at SOI. 
The islands are clearly visible in the middle of the channels from previous encounters. The core strand seems to be unperturbed at this stage of the orbit.
Fig. \ref{I10533fig6} shows the same numerical model, but Prometheus' mean anomaly
is $M=0.0^{\circ}$, which corresponds to Prometheus periapsis and F ring aposapsis. The streamers
from previous encounters are visible in the inner edge of the background sheet, and as expected streamers are more prominent to those observed in the SOI configuration. The gravitational perturbations in the core are easily seen in this plot. 

\section{Summary}
The anti-alignment configuration between Prometheus and the F ring is studied here, with special emphasis in the aspect of the channels. The approximated date for such an  event is November 21, 2009.
At this configuration the orbit of Prometheus (which is the inner object) is at its apoapsis while the ring (the outer object) is at its periapsis, and so their distance is at its minimum.

We found that the channels for this configuration go deeper in the ring and are wider compared to the ones reported for SOI in~\cite{Nature_2005}. We discovered a new feature which may be relevant to the F ring's structure: We called it the "island" since it resembles an islet in the middle of the channel. Every time Prometheus gets close to apoapsis ($\pm30^{\circ}$ from it), this new feature appears. This is a prediction of our model and should be visible when the system is in anti-alignment.
The number of collisions between Prometheus and the F ring particles was monitored.
Even for this configuration the number of collisions is quite low, only 0.038\% of the particles collided with
Prometheus per orbit. Thus these features are created by the gravitational perturbations of Prometheus and not by collisions, confirming~\cite{Nature_2005} results for the case of anti-alignment .
The conclusion is that the particles are only temporarily pulled away and then drift back 
into the ring one orbital period of Prometheus' later. The Cassini spacecraft should be able to see these predicted features in its extended mission that ends in July 1, 2010.

\section{Future work}
The detailed physics behind the channel's formation needs to be studied. Similarly,  the width of the channels as a function of the time needs to be explored. A comprehensive study of the collisions between Prometheus and the F ring particles needs to be done as well.

\ack
I would like to thank to C. D. Murray for his helpful suggestions and criticisms. Additionally, I would like to thank to M. W. Evans and Kevin Beurle for instructive discussions. I would like thank to my reviewers A. S. Bosh and S. Charnoz for their helpful comments and suggestions.
I thank Luis Aguilar for giving me the chance to develope this research and for all his comments and corrections. I thank David M. Clark for his comments and corrections. Lastly, I am grateful to IA-UNAM for their financial support.
\label{lastpage}

\bibliography{bibliography.bib}

%% Use the plainnat style for ``Icarus'' mode to display DOI numbers
%% among other things.  However, revert to the Elsevier elsart-harv
%% mode for ``Elsevier'' mode.
\bibliographystyle{plainnat}
% \bibliographystyle{elsart-harv}

%% --Tables-- 

\clearpage	% Make sure things don't run together.

\begin{table}[h]
\caption[Initial Orbital Elements for Prometheus and the F ring at the anti-alignment configuration]{
Initial Orbital Elements for Prometheus and the F ring at the approximated date for the anti-alignment configuration, corresponding to UT 4:54:24 on November 21, 2009. These were acquired using precessing ellipses, and the orbital elements reported in~\cite{Jacobson_2008} for Prometheus and ~\cite{Bosh_2002} for the F ring.
}
\label{I10533table}
\label{lasttable}	
\begin{tabular}{|c|c|c|}
\hline
Elements & Prometheus & F ring \\
\hline
$a (km)$ & $139380.0$ & $140223.7$ \\
\hline
$e$ & $0.0022$ & $0.00254$\\
\hline
$i(^{\circ})$ & $0.008$ & $0.0065$\\
\hline
$\widetilde{\omega}(^{\circ})$ & $236.2021$ & $56.0669$\\
\hline
$\Omega (^{\circ})$  & $117.9717$  & $29.2794$\\
\hline
$\lambda(^{\circ})$ & $77.5482$ & $-$\\
\hline
$ n (^{\circ}/day)$ & $587.2852370$ & $-$\\
\hline
\end{tabular}
\end{table}

\end{document}